\documentclass[%
 reprint,
 amsmath,amssymb,
 aps,
]{revtex4-1}
\usepackage{graphicx}
\usepackage{dcolumn}
\usepackage{bm} 
\usepackage{hyperref}
\usepackage{color}

\begin{document}
\preprint{APS/123-QED}
\title{Chemically non-perturbing SERS detection of catalytic reaction with black silicon}

\author{E. Mitsai$^1$, A. Kuchmizhak$^{1,2,*}$, E. Pustovalov$^{2}$, A. Sergeev$^{1,2}$, A. Mironenko$^{3}$, S. Bratskaya$^{2,3}$, D. P. Linklater$^{4,5}$, A. Bal\v{c}ytis$^{4,5}$, E. Ivanova$^{4}$, S. Juodkazis$^{4,5}$}
\email{alex.iacp.dvo@mail.ru}
\affiliation{$^1$Institute of Automation and Control Processes, Far Eastern Branch, Russian Academy of Sciences, Vladivostok 690041, Russia}
\affiliation{$^2$School of Natural Sciences, Far Eastern Federal University, Vladivostok, Russia}
\affiliation{$^3$Institute of Chemistry, Far Eastern Branch, Russian Academy of Sciences, Vladivostok 690041, Russia}
\affiliation{$^4$ Swinburne University of Technology, John st., Hawthorn 3122, Victoria, Australia}
\affiliation{$^5$Melbourne Centre for Nanofabrication, ANFF, 151 Wellington Road, Clayton, VIC 3168, Australia}

\date{\today}
\begin{abstract}
All-dielectric resonant micro- and nano-structures made of the high-index dielectrics recently emerge as a promising surface-enhanced Raman scattering (SERS) platform which can complement or potentially replace the metal-based counterparts in routine sensing measurements. These unique structures combine the highly-tunable optical response and high field enhancement with the non-invasiveness, i.e., chemically non-perturbing the analyte, simple chemical modification and recyclability. Meanwhile, the commercially competitive fabrication technologies for mass production of such structures are still missing. Here, we attest a chemically inert black silicon (b-Si) substrate consisting of randomly-arranged spiky Mie resonators for a true \textit{non-invasive} (chemically non-perturbing) SERS identification of the molecular fingerprints at low concentrations. Based on comparative in-situ SERS tracking of the para-aminothiophenol (PATP)-to-4,4` dimercaptoazobenzene (DMAB) catalytic conversion on the bare and metal-coated b-Si, we justify applicability of the metal-free b-Si for the ultra-sensitive non-invasive SERS detection at concentration level as low as 10$^{-6}$~M. We perform supporting finite-difference time-domain (FDTD) calculations to reveal the electromagnetic enhancement provided by an isolated spiky Si resonators in the visible spectral range. Additional comparative SERS studies of the PATP-to-DMAB conversion performed with a chemically active bare black copper oxide (b-CuO) substrate as well as SERS detection of the slow daylight-driven PATP-to-DAMP catalytic conversion in the aqueous methanol solution loaded with colloidal silver nanoparticles (Ag  NPs) confirm the non-invasive SERS performance of the all-dielectric crystalline b-Si substrate. Proposed SERS substrate can be fabricated using easy-to-implement scalable technology of plasma etching amenable on substrate areas over $10\times 10$~cm$^2$ making such inexpensive all-dielectric substrates promising for routine SERS applications, where the non-invasiveness is of mandatory importance.
\begin{description}
\item[PACS numbers]
\end{description}
\end{abstract}
\pacs{Valid PACS appear here}
\maketitle

\section{\label{sec:level1}Introduction}

For decades, the surface-enhanced Raman scattering (SERS) effect mainly based on the augmentation of the inelastic scattering of the probed molecules placed in the vicinity of the electromagnetic ``hot spots'' is considered as a non-invasive characterization technique allowing label-free quantitative identification of the vibrational molecular fingerprints, even at single-molecule level~\cite{Fleischmann74,Xu99,Kneipp99,Anker08,Kneipp}. To create such hot spots, the plasmon-active nanostructured substrates are widely used, allowing multi-fold enhancement of the Raman yield via electromagnetic effect as well as additional enhancement via the modification of the molecule electronic structure (a charge transfer).

Fast development in design of the SERS-active substrates have led to the growing interest in their applications for in-situ studies of the catalytic reactions allowing identification and quantification of the intermediates, transition states and final products~\cite{Harvey15}. Aside from such challenges as the chemical and thermal stability of the plasmon-active SERS substrates, their inert (non-invasive) character has been recently put into question after the demonstration of the spontaneous conversion of para-aminothiophenol (PATP) to 4,4`-dimercaptoazobenzene (DMAB) on the silver- and gold-coated substrates during SERS measurements~\cite{Huang10,Sun12,Wang14ACS,Zhang18}. Multiple efforts were undertaken to study the PATP, particularly owing to its affinity to the most-common plasmonic-active metals as well as appearance of the several characteristic Raman bands in its SERS spectra~\cite{Osawa94,Zhou06,Zhou07,Kim11,Ye12,Sun12,Dai15}. These bands were initially assigned to the b$_2$-type non-totally symmetric vibrations of the PATP molecule emerging in the detected spectra via the charge transfer (or chemical enhancement) mechanism~\cite{Osawa94,Kim11,Ye12}. However, more recent studies assigned those bands to the dimercaptoazobenzene (DMAB) molecule produced from PATP via a plasmon-driven catalytic oxidation~\cite{Sun12}. This shows that for  certain molecules, SERS is not always a non-invasive technique veiling the exact information on the vibrational fingerprints of the initial molecules deposited onto the plasmon-active SERS substrate. In this respect, the search for the chemically inert SERS substrates which can overcome this limitation allowing for a true non-invasive bio-identification with good sensitivity is of mandatory importance.

Recently, all-dielectric resonant micro- and nanostructures made of the high-index dielectrics (Si, Ge, GaAs, etc.) emerge as a promising platform which can complement or potentially replace the metal-based counterparts in routine SERS measurements \cite{Huang15,Alessandri,Krasnok17,Makarov17}. The main advantages of such structures are their low invasiveness, reproducibility, tunability of optical response, recyclability, as well as wide feasibilities for surface functionalization via covalent binding of selective receptors to provide host-guest molecular recognition \cite{Lin17,Bontempi17-2}. Generally, the isolated or ordered arrays of the high-index resonant structures supporting various types of Mie resonances were suggested for SERS measurements \cite{Caldarola15,Milichko17LPR,Cambiasso17,Dmitriev16}. Noteworthy, fabrication of such structures is challenging as it involves costly, non-scalable and time-consuming electron- or ion-beam milling techniques making all-dielectric resonators too expensive for routine biosensing experiments. Laser-induced forward transfer technique allows rapid fabrication of all-dielectric resonant structures via pulse-laser melting of the donor substrate resulting in ejection of the spherical-shape drops which can be transfered onto a receiving substrate \cite{Chichkov:2014,Dmitriev16-2}. Meanwhile, such laser-assisted fabrication of large-scale commercially competitive substrate still requires rather long fabrication cycles and complicated alignment (adjustment) procedures, while the shape of the produced resonant structures can be controllably tuned only in a rather narrow range of parameters.

In this study, we demonstrate an application of a chemically inert black-Si (b-Si) substrate consisting of randomly-arranged spiky Mie resonators for a true \textit{non-invasive} SERS identification of the molecular fingerprints at low concentrations. Based on comparative in-situ SERS tracking of the PATP-to-DMAB catalytic conversion 
on the bare and metal-coated b-Si, we justify applicability of the metal-free b-Si for the ultra-sensitive non-invasive SERS detection at concentration level as low as 10$^{-6}$. Supporting finite-difference time-domain (FDTD) calculations were undertaken to show the origin of the electromagnetic enhancement of the isolated spiky Si resonators in the visible spectral range.
By detecting the PATP-to-DMAB conversion on a chemically active bare black copper oxide (b-CuO) substrate as well as a slow-rate conversion photo-catalyzed by colloidal silver in an aqueous methanol solution on b-Si we further confirm the chemically non-perturbing character of SERS performance of the all-dielectric crystalline substrate.

\section{\label{sec:level1}Methods}
\textbf{Substrate fabrication.} The b-Si substrates were fabricated following a simple reactive ion etching procedure in O$_2$ and SF$_6$ gas mixture~\cite{13nc2838,15oe127103,16aplp076104}.
The b-CuO substrates are made by chemical oxidation of industrial grade copper foil (Flewsolutions Pty) using earlier developed processing steps~\cite{15n18299}.

\begin{figure*}[t]
\includegraphics[width=0.85\textwidth]{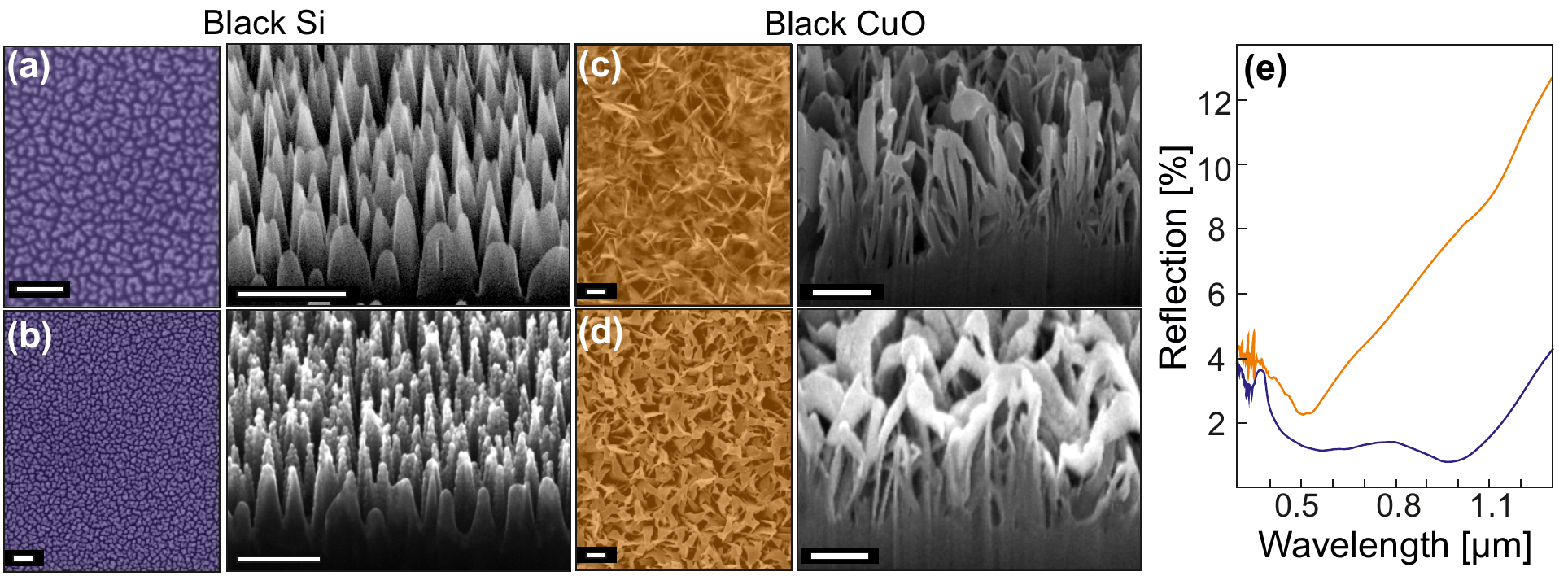}
\caption{\textbf{B-Si and b-CuO substrates and their optical properties.} (a-d) False-color normal-view SEM images of the bare (a) and the Ag-coated b-Si (b), as well as the bare (c) and Ag-coated black b-CuO (d). Scale bar corresponds to 500~nm. Side-view SEM images (in the right) show cross-sectional FIB cuts of the corresponding substrates. (e) Reflection coefficient of the bare b-Si (purple) and b-CuO substrates (orange).}
\label{fig:1}
\end{figure*}

Finally, gold and silver films of variable thicknesses ranging from 20 to 200~nm are coated onto the b-Si and b-CuO substrates using e-beam evaporation procedure. All films are deposited at a constant deposition rate of 1~nm/s while rotating the sample holder to ensure uniform deposition. SERS measurements on the  silver-coated substrates were performed within 24~hours after evaporation to avoid silver oxidation.

\textbf{Substrate characterization.} The nano-topography of the resulting bare and metal-coated blackend substrates are carefully inspected by the scanning electron microscopy (SEM, Carl Zeiss Ultra55+). Additionally, cross-section cuts of the bare and metal-capped blackend surfaces of both types are prepared to check the vertical size of the surface structures. To do this, the focused ion-beam milling (FIB, Carl Zeiss CrossBeam 1540) is utilized with a Ga$^+$-ion beam at 30~kV and current ranging from 50 to 200~pA. Higher ion-beam currents are applied to produce the initial cut, while a high-frequency beam scanning at low current was utilized to minimize the curtain effect on the cross sectional cut.

Reflection coefficient of both, the bare blackend substrates in the visible and near IR spectral ranges is measured using an optical integrating sphere spectrometer (Varian, Cary 5000).

\textbf{FDTD calculations.} Enhancement and local structure of the electromagnetic fields near the isolated tips of variable size on the b-Si surface are accessed using finite-difference time-domain modeling (FDTD, Lumerical Solutions package). The modeled geometry is extracted from the corresponding SEM images of the produced FIB cuts. The dielectric function of Si is modeled using the experimental library data~\cite{Palik}. The size of the square unit cell is $0.5\times 0.5\times0.5$~nm$^3$, while the computational volume is limited by perfectly matched boundaries. The linearly-polarized Gaussian light source with the lateral size fitting the size of the experimental optical spot is used to excite the structure. The total-field scattered field source of the same size is applied for calculations of the scattering spectra  (for details, see supporting material; Fig. S2).

\textbf{Raman spectroscopy.} A commercial Raman apparatus (Alpha, WiTec) equipped with a 532-nm CW semiconductor laser is used for in-situ Raman detection of the PATP-to-DMAB conversion. The linearly-polarized radiation is focused by an objective lens (100x, Carl Zeiss) with a numerical aperture ($NA$) of 0.9, providing the excitation of the local section ($\approx$0.41~$\mu$m$^2$) of the surface. The focal depth of the lens, which can be approximately estimated as a doubled Rayleigh length in air $\approx$$\lambda$(NA$^{-2}$)=1.3~$\mu$m ensures the uniform irradiation of the high-aspect-ratio textures of the used blackend surfaces. The incident power is tuned using a built-in attenuator. The Raman signal from the adsorbed molecules within the focal volume is collected with the same lens and analyzed using a grating-type spectrometer (1800~lines/mm) equipped with an electrically-cooled CCD camera.

In this study, the PATP is used as a model adsorbate owing to its strong SERS response, simple structure and affinity to the most common plasmonic materials as silver and gold used in this study. Moreover, these molecules represent the perfect probe to study the enhancement mechanisms in SERS. The drops of the alcoholic solution of the PATP at its molar concentration of 10$^{-6}$~M are deposited onto the surface of the SERS substrates using the home-build drop deposition system and Raman measurements are conducted after complete evaporation of the drop.

Additionally, the bare or silver-capped b-Si substrates are attested for \emph{in situ} detection of the PATP-to-DMAB conversion in water/methanol solution (1/1, v/v) at the PATP molar concentration of 5$\cdot$10$^{-2}$M. In this experiment, SERS substrates are immersed into a quartz cell filled with the above mentioned solution, where slow PATP-to-DMAB conversion is stimulated via surface plasmon resonance (SPR)-assisted catalysis by colloidal silver nanoparticles (Ag NPs) at PATP/Ag molar ratio 500/1. Spherical-shape Ag NPs having the diameter ranging from 5 to 10~nm are synthesized via citrate reduction method and purified by dialysis (see details in the Supporting information). For SERS studies, the laser radiation is focused onto the substrate surface through the solution with long-working-distance lens of $NA = 0.25$ exciting the local section of $\approx$1.3~$\mu$m$^2$) of the b-Si surface, while similar lens and spectrometer are used to analyze the Raman response. The position of the focal spot is calibrated by maximizing yield from the Raman band of the crystalline Si. The sealed quartz cells with the substrate and solutions are maintained under daylight exposure between the SERS measurements.

\section{\label{sec:level1}Results and Discussions}

\begin{figure*}[t]
\includegraphics[width=0.75\textwidth]{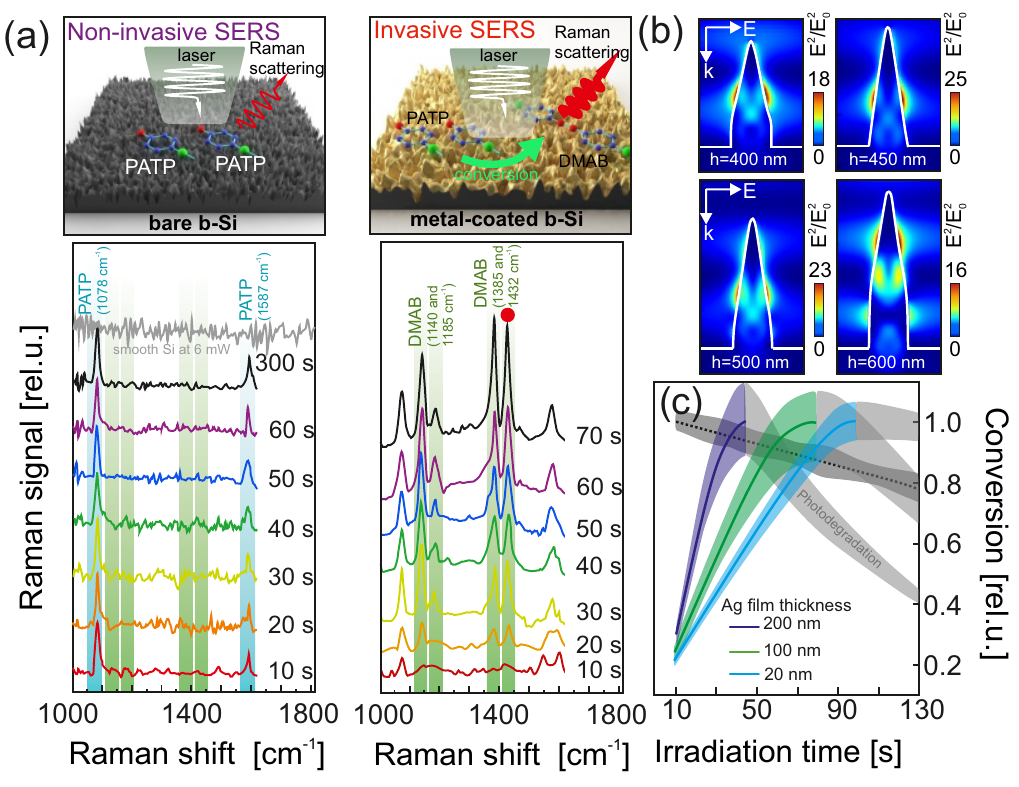}
\caption{\textbf{SERS detection of the PATP-to-DMAB conversion with bare and silver-coated b-Si.} (a) Series of Raman  SERS spectra of the PATP layer adsorbed on the bare (left) and Ag-coated (right) b-Si substrates. 
The total irradiation time of the PATP molecules is indicated near each spectrum, while the accumulation time for each spectrum is 10~s. The characteristic bands of the PATP at 1078 and 1587~cm$^{-1}$ are highlighted by the blue-color areas, while four specific bands attributed to DMAB are highlighted by green. The Raman band at 1432 $cm^{-1}$ used to access the conversion are marked with the red circle. Each spectrum is shifted on the vertical axis for better displaying. (b) Squared electric-field amplitude inside and near the isolated spiky Si resonators of variable height ranging  from 400 to 600 nm calculated at 532~nm. (c) PATP-to-DMAB conversion versus laser irradiation time. The conversion is defined as a normalized intensity of the DMAB characteristic Raman band at 1432~cm$^{-1}$ measured for the b-Si coated with 20- (green), 100- (green) and 200-nm thick (purple) Ag film (blue) at 13.1~$\mu$W/$\mu$m$^2$. Similar conversion measured at 120~$\mu$W/$\mu$m$^2$ for b-Si coated with 20-nm-thick Ag film is shown by the dashed curve. Grey-color area indicate the photo-degradation of the analyte.}
\label{fig:1}
\end{figure*}

As mentioned, the surface of the b-Si contains randomly-arranged vertically oriented spikes with the height of about 600$\pm$150~nm (Fig.1(a)), tip curvature radius smaller than 10~nm and an average density of $\approx$60 spikes/$\mu$m$^2$. Non-uniform coverage of the b-Si with the deposited Ag (or Au) film produces even more textured surface with strong plasmonic response and high SERS yield (Fig.~1(b)). In a sharp contrast, b-CuO surface has substantially more chaotic orientation of the textures with respect to the surface normal, while their average height exceeds 1 $\mu$m with the broad deviation of sizes in a lateral dimension (Fig.~1(c)). The deposition of the metal film results in formation of the metal flakes on the textured surface (Fig.~1(d)). Noteworthy, the density of the spikes is similar to those for b-Si, while the SERS yield is also comparable, as shown in previous studies~\cite{Schmidt12,Gervinskas13,15n18299}.
Meanwhile, the density of the surface textures both substrate ensure that several ($\approx$10-20) randomly arranged spiky nano-features will be always within the laser irradiated spot (see Methods for details). Such dense surface structure allows both substrates to almost perfectly absorb the radiation in the visible spectral range as it is illustrated by the measurements of the reflection coefficient (Fig.~1(e)). Specifically, for the 532-nm laser wavelength used further for Raman spectroscopy, the reflection coefficient as small as 2.1\% is found for the bare b-CuO, dropping down to 1.3\% - for the black Si. The nano-texture is acting as a gradient refractive index region and acts as an anti-reflection coating.

For the probed spectral region defined by the spectrometer grating, only two distinct Raman bands at 1587 and 1078~cm$^{-1}$ (marked in blue; Fig.2(a, left)) which are the $\nu$CS(a1) and $\nu$CC(a1) vibration modes of the PATP molecules~\cite{Cao05,Osawa94} capping the surface of the bare b-Si are distinguishable. Time-resolved control SERS measurements show that no substantial changes of the relative intensity of the both identified modes occur within several minutes of measurements (Fig.~2(a,left)). Noteworthy, even a fifty-fold increase of the irradiation intensity or longer accumulation time do not allow detection of the PATP on the surface of the smooth crystalline Si substrate (the upper-most curve in Fig.~2(a, left)), indicating the pronounced SERS effect for the bare b-Si. For the both PATP Raman bands, the non-metallic b-Si substrate provides an averaged enhancement factor of $\approx$10$^3$ as compared to the smooth crystalline Si surface. This is larger by more than an order of magnitude considering the effect of an increased surface area.

Particularly, this enhancement can be attributed to the more efficient absorption of laser radiation (decreased reflectivity) as well as to the enhanced electromagnetic fields near the cone-shaped Mie resonators randomly distributed across the b-Si surface. Supporting FDTD calculations are performed to access the enhancement of the electromagnetic (EM) field near the isolated b-Si spiky resonators of variable sizes (see Methods for details). The calculated enhancement of the normalized squared field amplitude (E)$^2$/(E$_0)^2$ as high as 25 is found on the surface of metal-free spiky Si resonators, sufficiently high to boost the SERS yield, which is proportional to the forth power of the EM amplitude, via the field enhancement mechanism (see Fig.~2(b)). Indeed, the position of the EM hot spots on the surface of the spiky tip as well as the maximal amplitude are size-dependent (Fig.~2(b)). Meanwhile, for the produced b-Si with an averaged height of 550$\pm$150~nm, the squared EM-field amplitude never drops below 15. Noteworthy, according to our calculations the field enhancement inside the tip reaches (E)$^2$/(E$_0)^2$$\approx$12 for the certain tipped structures, indicating that each isolated structure can act as a typical all-dielectric resonator~\cite{Kuznetsov16, Bontempi17,Krasnok17}, which can support various types of modes (see the Fig.~S2 in the Supporting information). In this respect, the b-Si substrate can be considered as a disordered array of the densely packed spiky Mie resonators. As the distance between the neighboring structures is rather small (see Fig.~1(a,b)), the enhancement of the EM field in their gaps can also boost the SERS yield. High density of the structures potentially provide broadband excitation in the visible spectral range ensuring that for the fixed size deviation of the Si resonators on the surface, at least several of them will be resonantly excited by the visible laser radiation, even under tight focusing conditions.

In addition to the substantially enhanced EM field near the surface of such dielectric resonators, use of crystalline materials gives an additional modality to the chemical SERS measurements allowing in situ detection of the local temperature distribution via tracking the spectral shape and position of the temperature-dependent c-Si Raman band~\cite{Balkanski83,Burke93}, which can be resonantly excited within the Si resonator~\cite{Dmitriev16,Aouassa17}. In the range of applied intensities used for Raman measurements, we did not observe any substantial variations of the shape and position of the c-Si band (see Fig.~S3 in the Supporting information) indicating potentially weak contribution of the temperature effect to the obtained SERS data. This is consistent with a recently reported laser trapping/pinning of micro-beads on bare b-Si at high light intensities which usually creates strong temperature gradients and convection flows on plasmonic metal coated substrates \cite{Shoji17}.

\begin{figure}[t]
\includegraphics[width=0.95\linewidth]{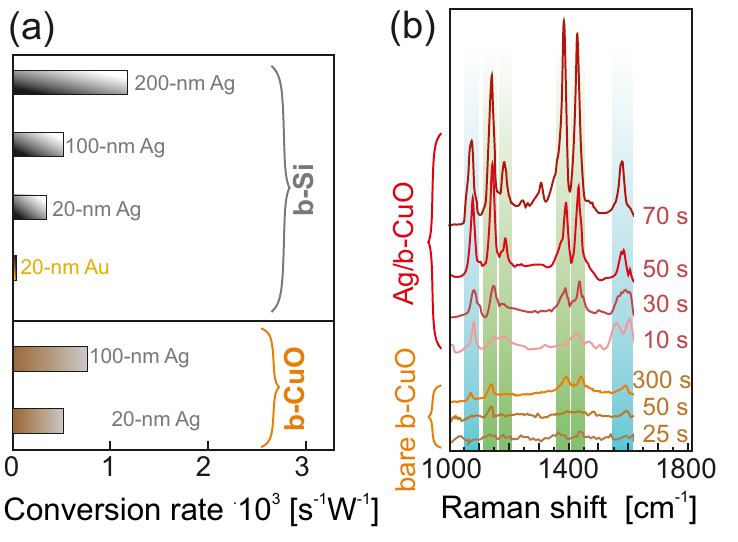}
\caption{(a) Estimated PATP-to-DMAB conversion rate for Ag- and Au-coated b-Si and b-CuO substrates. (b) Series of SERS spectra of the PATP layer adsorbed on the bare and Ag-coated b-CuO substrates.}
\label{fig:3}
\end{figure}

In a sharp contrast, for b-Si coated with a 100-nm thick Ag film (Ag/b-Si) several evident Raman bands at 1140, 1185, 1385 and 1432~cm$^{-1}$ appear in the measured spectra (green-color areas in Fig.~2(a, right)). Time-resolved SERS measurements show that after first 10~seconds of laser irradiation of molecules on the substrate, only the low-intensity bands assigned to PATP can be identified. The appearance and continuous increase in intensity of all the bands attributed to the DMAB can be seen for longer irradiation times presumably indicating the monotonous conversion of the PATP to DMAB via a plasmon-driven catalytic reaction. 
It should be stressed that the  bare b-Si substrate expectedly provides about 2 orders of magnitude lower SERS yield for the PATP bands comparing to the metal-coated one. This presumably is due to the absence of the chemical SERS enhancement mechanisms proceeding via the chemical interactions at the substrate-analyte interface. Meanwhile, for such metal-free chemically inert substrate, SERS identification of the adsorbed analyte molecules has true non-invasive character.

\begin{figure*}[t]
\includegraphics[width=0.7\textwidth]{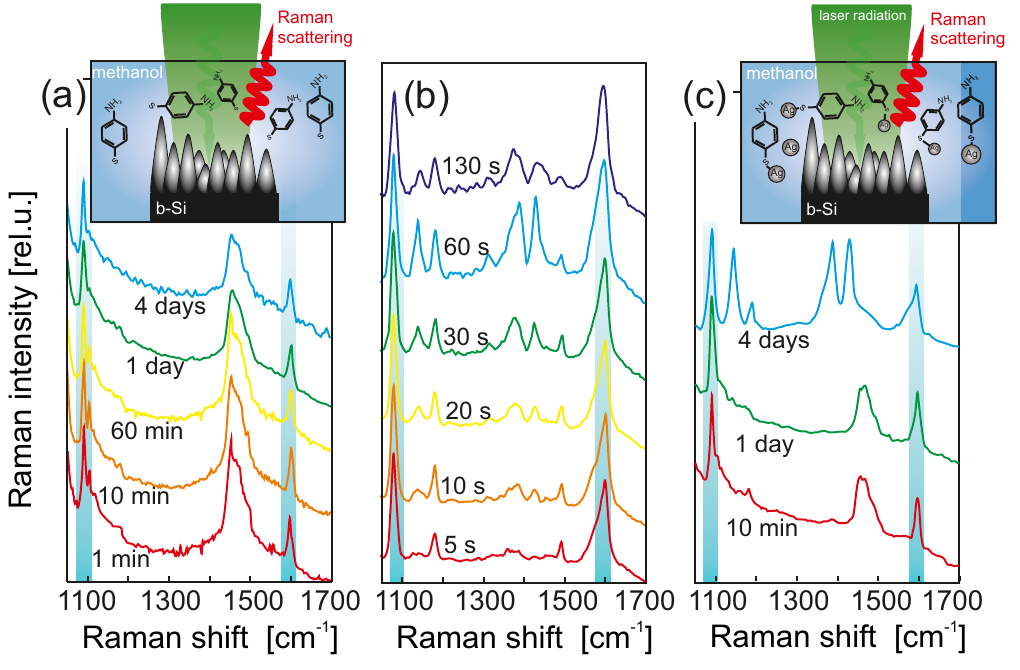}
\caption{\textbf{Non-invasive SERS detection of the PATP-to-DMAB daylight-driven catalytic conversion in the aqueous methanol solution.} Time-resolved SERS spectra of the PATP at molar concentration of 5$\times$10$^{-2}$ dissolved in pure (a,b) and Ag NP-loaded aqueous methanol solution (c). For these measurements, bare (a,c) and Ag-coated (b) b-Si substrates were used. Detection procedure is schematically illustrated in the insets. The accumulation times for each spectrum are 30 (a,c) and 5~s (b). For measurements with bare b-Si substrates, both solutions were maintained under daylight irradiation, while the total incubation time is indicated near each spectrum.}
\label{fig:3}
\end{figure*}

The relative intensity ratio of the Raman bands at 1432 and 1587~cm$^{-1}$ is typically used to access the PATP-to-DMAB conversion rate~\cite{Kim11,Dai15}. However, both the monomer and dimer molecules contribute to the total intensity of the Raman band at 1587~cm$^{-1}$ complicating the analysis of the conversion kinetics and total yield.
In this study, without loss of generality, we use the time evolution of the normalized intensity of the DMAB Raman band at 1432~cm$^{-1}$ to preliminary access the PATP-to-DMAB conversion, assuming that the concentration of DMAB increases to a certain maximum value within the laser-exposed volume for the used substrate. Figure~2(c) shows the average conversion plotted against the laser irradiation time of the PATP molecules deposited onto the Ag/b-Si substrate for various thicknesses of metal coating and irradiation intensities of 13.1 and 120~$\mu$W/$\mu$m$^2$. At low intensity level, the monotonous growth of the DMAB signal is observed for all Ag/b-Si substrates within the first 50-90 seconds of the irradiation cycle (colored solid curves in Fig.~2(c)). Upon reaching maximum, the DMAB Raman band intensity decreases due to photo-degradation of the analyte and a growing contribution of luminescence background from newly formed complexes. This process is illustrated as the grey-color areas in Fig.~2(c). The transition between two regimes indicates the end of the PATP-to-DMAB conversion within the exposed volume having a fixed number of the initial PATP molecules.
Meanwhile, almost 10-fold higher intensity of 120~$\mu$W/$\mu$m$^2$ of the laser radiation results in complete conversion of the PATP-to-DMAB within the first 10 seconds of the laser exposure (dashed curves in Fig.2(c)).
Noteworthy, the higher Raman yield for the characteristic PATP (DMAB) bands, as well as faster reaction and degradation rates correspond to thicker Ag films, indicating the key role of metal in the SERS enhancement, PATP-to-DMAB conversion and photo-degradation processes~\cite{Wu17}.


The obtained results are further analyzed by fitting the initial part of the intensity evolution of the DMAB Raman band at 1432~cm$^{-1}$ (see Fig.~2(c)) and normalizing the obtained line slope on the incident laser intensity. In this way, we estimated the average PATP-to-DMAB conversion rate for various types of the applied film thickness, summarizing the obtained trends in Fig.~3(a). Noteworthy, SERS enhancement of PATP (DMAB) bands measured on the Au-coated b-Si is almost order of magnitude weaker, which can be explained in terms of more favorable EM enhancement under 532-nm excitation of the Ag-coated b-Si. Additionally, the chemically active Ag coating provides faster PATP-to-DMAB conversion compared to the Au coating of the same thickness (see also Fig.~S4 in Supporting information)~\cite{Zhang17}. Aside from higher reactivity of the silver toward the hydrogen sulfide~\cite{Mironenko16} and its derivatives, responsible for easier formation of a charge transfer complex and chemical amplification, it have been reported that the formation of DMAB on the Ag surfaces is much easier under the same experimental condition than that on Au ones, considering an easier formation and a higher activity of triplet oxygen molecules ($^3$O$_2$) on the Ag surfaces~\cite{Huang14,Xua15}. Furthermore, with the activation of $^3$O$_2$ species driven by plasmon resonances, the 514.5~nm excitation was shown to be more efficient for DMAB formation on the Ag surfaces, whereas, the 632.8~nm excitation was preferred on the Au ones~\cite{Xua15}.


In a sharp contrast to the chemically stable bare b-Si, the b-CuO which surface can contain the chemically active metallic sites, as it was confirmed by previous XPS analysis~\cite{15n18299}. This would allow the PATP molecules to be converted to the DMAB under laser exposure even without any metal coating. It should be stressed, the copper oxide have been shown to be a catalyst for the amine oxidation~\cite{Al-Hmoud}. In our experiments, for bare b-CuO we found rather slow PATP-to-DMAB conversion rate (see three bottom curves in Fig.~3(b)), while the coverage of the b-CuO with the 100-nm thick Ag film provides the pronounced SERS yield with an average conversion rate of 750~(s$\times$W)$^{-1}$ (four upper spectra in Fig.~3(b)).

Although, the bare b-CuO demonstrates slightly better SERS performance being compared to the bare b-Si substrate, the latter provides the way for non-invasive SERS registration with temperature feedback modality and moderate SERS yield allowing non-perturbing identification of the molecule fingerprints at relatively low concentration. To further illustrate this remarkable feature of the bare b-Si, we studied its applicability for non-invasive detection of the slow daylight-driven PATP-to-DAMP catalytic conversion in the aqueous methanol solution loaded with the Ag NPs (see Methods for details).

For pure solution of the PATP (without Ag NPs), corresponding time-resolved SERS spectra detected near the b-Si surface demonstrate only two Raman bands attributed to the monomer molecules even within rather long period of 4 days under daylight exposure (Fig.~4(a)). It should be noted, for the same solution silver-coated (100-nm thick) b-Si converts the PATP-to-DMAB within the first minute of irradiation maintaining the conversion rate similar to those obtained for the dried layer (see Fig.~S5 in the Supporting information). This makes the metal-coated substrate non-applicable for SERS detection of the considered model reaction in solutions, as the sensing surface is covered by the PATP layer owing to strong affinity of the PATP to the plasmon-active metals. Such SERS measurements obtained from this near-surface PATP layer veil the exact information on the slow conversion kinetics proceeding in the solution.

Finally, to stimulate the slow daylight-driven PATP-to-DMAB catalytic conversion, the Ag catalyst was added to the PATP solution and the same time-resolved measurements were performed on the bare b-Si substrate immersed into the solution (see Methods for details). These measurements show that only after 4 days of daylight exposure a clear fingerprint of DMAB molecules appeared in the measured SERS spectra (Fig.~4(b)). Hence, the bare b-Si can be considered as a substrate allowing chemically non-perturbing study of the catalytic kinetics in the solution.

\section{\label{sec:level1}Conclusions and outlook}
To conclude, we successfully demonstrate the non-invasive SERS identification of the PATP molecules with non-metallic b-Si surface at concentration as low as 10$^{-6}$ M and an average SERS enhancement of 10$^3$. The enhancement can be attributed to the efficient absorption of the laser radiation by anti-reflective b-Si as well as enhancement of the electromagnetic field near the spiky nano-features acting as Mie resonators, as it was confirmed by the supporting FDTD calculations. Thermal effects can be monitored by observation of 521~cm$^{-1}$ crystalline Si vibration mode and adds additional value for precise monitoring of (photo-)catalytic reactions on b-Si substrates.
More importantly, comparative in situ SERS tracking of the PATP-to-DMAB catalytic conversion on the Ag- and Au-capped b-Si as well as bare and metal-coated b-CuO confirms the chemically non-perturbing character of the non-metallic all-dielectric crystalline substrate. This remarkable feature allows to use such substrates for non-invasive SERS tracing of the catalytic processes in the solution, which was demonstrated by detecting slow PATP-to-DMAB conversion catalyzed by the colloidal silver nanoparticles under daylight exposure in the aqueous methanol solution.

Mie resonances explored here for the chemically non-perturbing SERS detection can be also exploited in  agglomerated networks of nanoparticles which provided detection of atmospheric oxygen and nitrogen as well as polymer on the laser ablated surfaces (without metal coating) \cite{Bellouard17}. Here demonstrated, a single-point SERS measurement technique on dielectric surfaces can be combined with a surface mapping which provides statistical distribution of the hot-spots over the surface and can be directly linked to the analyte concentration~\cite{15jb567}. Quantitative detection of analytes in solution is a required key technology of bio-medical and environmental sensing.

{\begin{acknowledgments}
The work was partially supported by the FEBRAS Program for Basic Research ``Far East'' (projects Nos. 18-3-002, 18-3-012). A.K. acknowledges the partial support from Russian Foundation for Basic Research (17-02-00571-a) and RF Ministry of Science and Education (Contract No. MK-3287.2017.2) through the Grant of RF President. We are grateful to M. Larkins for b-CuO SERS test samples and discussions.
\end{acknowledgments}}

\end{document}